\documentclass[twocolumn,prl,showpacs,byrevtex]{revtex4}
\usepackage{epsf,graphicx,amssymb}

\begin{document}
\title{Wave turbulence on the surface of a ferrofluid submitted to a magnetic field}
\author{Fran\c cois Boyer}
\author{Eric Falcon}
\email[E-mail: ]{eric.falcon@univ-paris-diderot.fr}
\affiliation{Laboratoire Mati\`ere et Syst\`emes Complexes (MSC), Universit\'e Paris Diderot, CNRS (UMR 7057)\\10 rue A. Domon \& L. Duquet, 75 013 Paris, France}

\date{\today}

\begin{abstract}  
We report the observation of wave turbulence on the surface of a ferrofluid mechanically forced and submitted to a static normal magnetic field. We show that magnetic surface waves arise only above a critical field. The power spectrum of their amplitudes displays a frequency-power law leading to the observation of a magnetic wave turbulence regime which is experimentally shown to involve a 4-wave interaction process. The existence of the regimes of gravity, magnetic and capillary wave turbulence is reported in the phase space parameters as well as a triple point of coexistence of these three regimes. Most of these features are understood using dimensional analysis or the dispersion relation of the ferrohydrodynamics surface waves.

\end{abstract}
\pacs{47.35.Tv,47.65.Cb,47.27.-i}

\maketitle
Wave turbulence is an out-of-equilibrium state where waves interact with each other nonlinearly through $N$-wave resonance process. The archetype of wave turbulence is the random state of ocean surface waves, but it appears in various systems: capillary waves \cite{Wright96,Falcon07}, plasma waves in solar winds, atmospheric waves, optical waves, and elastic waves on thin plates \cite{ZakharovLivre}. Recent laboratory experiments of wave turbulence have shown new observations such as intermittency~\cite{Falcon07b}, fluctuations of the energy flux \cite{Falcon08}, and finite size effect of the system \cite{Falcon07,Denissenko07}. Some of these phenomenon have recently been considered theoretically \cite{Choi05}. Wave turbulence theory allows to analytically derive stationary solutions for the wave energy spectrum as a power law of frequency or wave number \cite{ZakharovLivre}. The spectrum exponent and the number $N$ of resonant waves depend on both the wave dispersion relation and the dominant nonlinear interaction. Several theoretical questions are open, notably about the validity domain of the theory \cite{Choi04}, and the possible existence of solutions for non dispersive systems \cite{Connaughton03}. In this context, finding an experimental system where the dispersion relation of the waves could be tuned by the operator should be of primary interest to test the wave turbulence theory.
 
A ferrofluid is a suspension of nanometric ferromagnetic particles diluted in a liquid displaying striking properties: the Rosensweig instability \cite{Cowley67}, the labyrinthine instability, magnetic levitation \cite{Rosen}. In contrast with usual liquids, the dispersion relation of surface waves on a ferrofluid displays a minimum which depends on the amplitude of the applied magnetic field \cite{Broaweys99,Embs}. Thus, one can easily tune the dispersion relation of surface wave from a dispersive to a non dispersive one with just one single control parameter. To our knowledge, no experimental observation of wave turbulence on a magnetic fluid has been reported. Here, we study the wave turbulence on the surface of a ferrofluid submitted to a normal magnetic field. We observe for the first time a regime of magnetic wave turbulence. We characterize this regime by measuring the power spectrum and distribution of the magnetic wave amplitude.
 
\begin{figure}[t!]
\centerline{
\epsfxsize=85mm
\epsffile{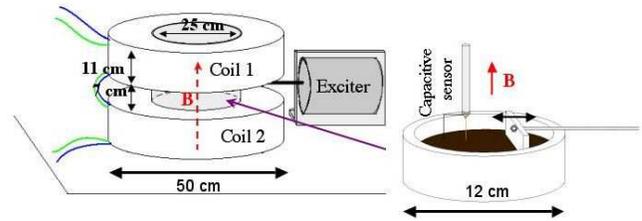} 
}
\caption{Experimental setup}
\label{fig01}
\end{figure}
 
The experimental setup is shown in Fig.\ \ref{fig01}. It consists of a cylindrical container, 12 cm in inner diameter and 4 cm in depth, filled with a ferrofluid up to a depth $h=2$ cm. The ferrofluid used is a ionic aqueous suspension synthesized with 8.5\% by volume of maghemite  particles (Fe$_2$0$_3$ ; 7.6 $\pm 0.36$ nm in diameter) \cite{Talbot}. The properties of this magnetic fluid are: density, $\rho=1324$ kg/m$^3$, surface tension, $\gamma=59\times 10^{-3}$ N/m, initial magnetic susceptibility, $\chi_i=0.69$, magnetic saturation $M_{sat}=16.9\times 10^{3}$ A/m, and estimated dynamic viscosity 1.2 $\times 10^{-3}$ Ns/m$^2$. The container is placed between two horizontal coaxial coils, 25 cm (resp. 50 cm) in inner (resp. in outer) diameter, 7 cm far apart.  A DC current is supplied to the coils in series by a power supply (50 V/35 A), the coils being cooled with water circulation. The vertical magnetic induction generated is 99\% homogeneous in the horizontal plane \cite{Broaweys99}, and is up to 780 G. It is measured by a Hall probe located in the center near the surface of the container. Surface waves are generated on the ferrofluid by the horizontal motion of a rectangular plunging Teflon wave maker driven by an electromagnetic vibration exciter. The wave maker is driven with low-frequency random vibrations (typically from 1 to 5 Hz). The amplitude of the surface waves $\eta(t)$ at a given location is measured by a capacitive wire gauge (plunging perpendicularly to the fluid at rest) with a 7.1 mm/V sensitivity  \cite{Falcon07}. $\eta(t)$ is recorded by means of an acquisition card with a 4 kHz sampling rate, low-pass filtered at 1 KHz during 300 s, leading to $1.2\times 10^{6}$ points recorded.

In the deep fluid approximation, the dispersion relation of linear inviscid surface waves on a magnetic fluid submitted to a magnetic induction $B$ perpendicular to its surface, reads \cite{Rosen}
\begin{equation}
\omega^2=gk-\frac{f[\chi]}{\rho\mu_0}B^2k^2+\frac{\gamma}{\rho}k^3
\label{rdtheo}
\end{equation}
where $\omega$ is the wave pulsation, $k$ its wavenumber, $g=9.81$ m/s$^2$ the acceleration of the gravity,  $\mu_0=4\pi \times 10^{-7}$ H/m the magnetic permeability of the vacuum, and $f[\chi]\equiv \chi^2/[(2+\chi)(1+\chi)]$. $\chi$ is the magnetic susceptibility of the ferrofluid which depends on the applied magnetic field, $H$, through Langevin's classical theory \cite{Abou97} 
\begin{equation}
\chi(H)=\frac{M_{sat}}{H}\mathcal{L}\left(\frac{3\chi_iH}{M_{sat}}\right)
\label{aimantation}
\end{equation}
where $\mathcal{L}(x)\equiv \coth{(x)}-1/x$, and thus on the magnetic induction, $B$, through an implicit equation since
 \begin{equation}
B=\mu_0(1+\chi)H
\label{BvsH}
\end{equation}
For $B=0$, the dispersion relation of Eq.\ (\ref{rdtheo}) is monotonous, and is dominated by the gravity waves at small $k$, and by the capillary waves at large $k$. When $B$ is increased, the quadratic term $-B^2k^2$ increases, and the dispersion relation becomes non monotonous: an inflection point appears at $B=0.93B_c$, then a minimum which leads, at $B=B_c$, to the Rosensweig instability \cite{Rosen}.  This stationary instability (a hexagonal pattern of peaks on the ferrofluid surface) occurs when $\omega^2(k)$ becomes negative, that is for a critical induction $B^2_c=2\mu_0\sqrt{\rho g \gamma}/f[\chi(Hc)]$ where $\chi(H_c)$ is determined using Eqs.\ (\ref{aimantation}) and (\ref{BvsH}) \cite{Cowley67}. This leads to the theoretical value of the critical  induction $B_c=292.3$ G for the Rosensweig instability of our ferrofluid. When $B$ is slowly increased, a direct visualization of the surface leads to an experimental value of $B_c=294\pm2$ G which is close to the above expected value.

\begin{figure}[ht!]
\centerline{
  \epsfxsize=75mm
  \epsffile{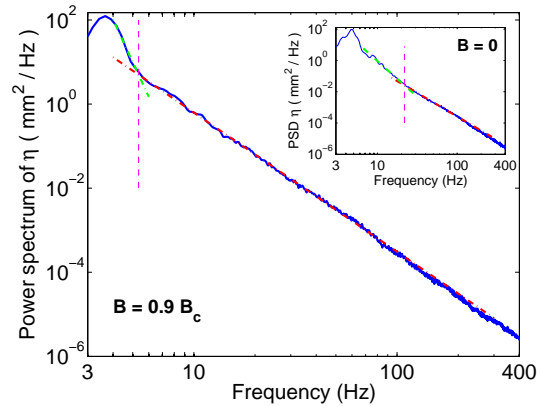}
}
\caption{Power spectrum of $\eta(t)$ for two values of $B$. Inset: $B=0$: Gravity and capillary wave turbulence regimes. Dashed lines have slopes -4.6 and -2.9. Crossover: $f_{gc} \simeq 20$ Hz. Main: $B=0.9B_c$: Magneto-capillary wave turbulence. Dashed line has slope -3.3. Crossover: $f_{gc} \simeq 5$ Hz. Forcing parameters: 1 - 5 Hz. $B_c= 294$ G.}
\label{fig02}
\end{figure}

The power spectrum of the wave amplitude on the surface of the  ferrofluid is shown in Fig.\ \ref{fig02} for different  applied magnetic induction $B$.  For $B=0$ (see the inset of Fig.\ \ref{fig02}), it displays similar results than those found with a usual fluid \cite{Falcon07}: two power laws corresponding to the gravity and capillary turbulence wave regimes. The capillary regime is found to scale as  $f^{-2.9\pm 0.1}$ in good agreement with the prediction of weak turbulence theory in $f^{-17/6}$ \cite{Zakharov67Cap}, and the gravity regime is found in  $f^{-4.6}$. The exponent of the gravity cascade being known to depend on the forcing parameter \cite{Falcon07,Denissenko07} (in contrast with the theoretical prediction $\sim$ $f^{-4}$ \cite{Zakharov67Grav}), it is thus only fitted to measure the crossover frequency $f_{gc}$ between gravity and capillary regimes.  As previously reported with usual fluids \cite{Falcon07}, $f_{gc}$ is also found here to decrease (from 26.6, 21 to $17.2 \pm 0.3$ Hz for a random forcing of 1 - 6 Hz, 1 - 5 Hz, and 1 - 4 Hz). The expected value is given by $f_{gc}=\frac{1}{\pi}\sqrt{2g/l_c}\simeq 15.2$ Hz where $l_c=\sqrt{\gamma/(\rho g)}$ \cite{Falcon07}. Consequently, in order to study the crossover frequency  dependence with the magnetic induction $B$, one has to rescale it by its value at $B=0$ such as $\tilde{f}(B)\equiv f_{gc}(0)\times f(B)/f(0)$ (see below).

For $B\neq0$, the power spectrum of wave amplitudes shows two striking new results. As shown in Fig.\ \ref{fig02}, the crossover frequency is strongly decreased down to roughly 5 Hz, and a power law in $f^{-3.3}$ appears in roughly all the accessible frequency range. Let us try first to understand this latter observation. The power spectrum of $\eta(t)$ can be derived  by dimensional analysis for the gravity and capillary wave turbulence regimes, respectively as \cite{Connaughton03} $S^{grav}_{\eta}(f) \sim \epsilon^{1/3}gf^{-4}$ and $S^{cap}_{\eta}(f) \sim \epsilon^{1/2}\left(\frac{\gamma}{\rho}\right)^{1/6}f^{-17/6}$, where $\epsilon$ is the energy flux [dimension $(L/T)^3$]. Since gravity wave turbulence is a 4-wave interaction process, and capillary waves a 3-wave one, it gives the dependence of the energy flux exponent in $\epsilon^{1/(N-1)}$ for a $N$-wave process \cite{Connaughton03}.
One can also derive dimensionally the power spectrum for the magnetic wave turbulence regime as
\begin{equation}
S^{mag}_{\eta}(f) \sim \epsilon^{\alpha}\left(\frac{B^2}{\rho\mu_0}\right)^{\frac{2-3\alpha}{2}}f^{-3} {\rm \ \ , }
\label{spectremag}
\end{equation}
In contrast with the above dispersive systems, this $-3$ frequency exponent does not depend on the energy flux exponent $\alpha$, that is on the number $N$ of resonant waves. The frequency exponent predictions for the magnetic and capillary regimes, respectively $-3$ and $-17/6\simeq -2.8$ can not be distinguished experimentally within our experimental accuracy. This could explain that only one single slope is observed on Fig.\ \ref{fig02} which thus corresponds to a ``magneto-capillary'' wave turbulence regime. 

\begin{figure}[t!]
\centerline{
\epsfxsize=75mm
\epsffile{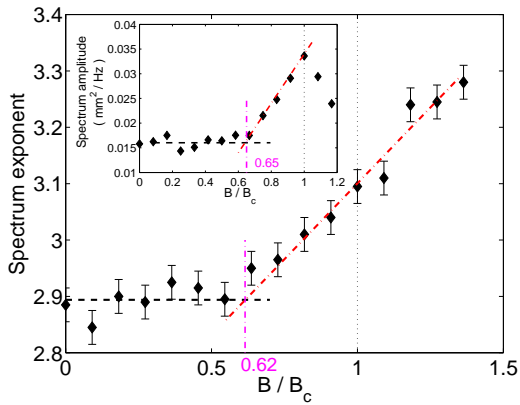} 
}
\caption{Exponent of the magneto-capillary spectrum as a function of the dimensionless magnetic induction $B / B_c$. Inset: Amplitude of the power spectrum (averaged between 8 and 18 Hz) as a function of $B / B_c$. Forcing parameter: $1 \leq f \leq 4$ Hz. $B_c= 294$ G.}
\label{fig03}
\end{figure}

Figure\ \ref{fig03} shows the frequency exponent of the magneto-capillary spectrum when $B$ is increased. For small $B$, capillary waves are dominant and the exponent is found roughly constant $\simeq -2.9$ in good agreement with the capillary prediction. When $B/B_c \geq 0.65$, magnetic waves becomes dominant (see below), and the exponent begins to slightly increase with $B$ up to -3.1 in rough agreement with the $-3$ prediction of Eq.\ (\ref{spectremag}). For $B/B_c\geq 1$, the Rosensweig instability occurs, and the spectrum exponent strongly changes with $B$. This could be attributed to the growing of the hexagonal pattern. The inset of Fig.\ \ref{fig03} shows the power spectrum amplitude (averaged between 8 and 18 Hz) as a function of $B$. The spectrum amplitude is found roughly constant when $B$ is increased till $B/B_c\simeq0.65$, onset of the magnetic waves. Above this value, the spectrum amplitude increases roughly linearly with $B$ up to $B/B_c=1$ where the instability occurs. Using Eq.\ (\ref{spectremag}), one can thus deduce from this linear dependence that $2-3\alpha=1$, and thus $\alpha=1/3$. This experimentally shows that the magnetic wave turbulence observed involves a 4-wave interaction process.

\begin{figure}[t!]
\centerline{
\epsfxsize=75mm
\epsffile{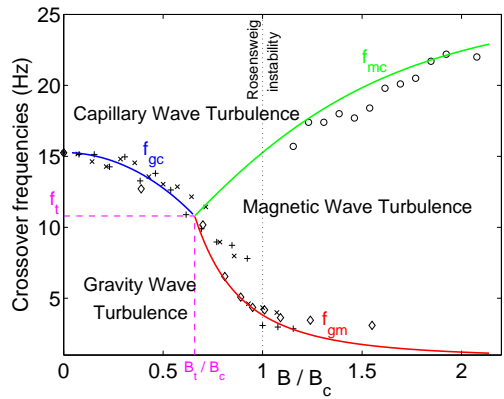} 
}
\caption{Rescaled crossover frequencies, $\tilde{f}(B)\equiv f_{gc}(0)\times f(B)/f(0)$ as a function of $B$ for different frequency bandwidths of the random forcing: ($\times$) 1 to 4 Hz, ($\diamond$) 1 to 5 Hz, and ($+$ or $\circ$) 1 to 6 Hz.  Theoretical curves $f_{gc}$, $f_{gm}$ and $f_{mc}$ are respectively from Eqs.\ (\ref{fgc}), (\ref{fgm}) and (\ref{fmc}). The triple point ($f_t=10.8$ Hz , $B_t/B_c=0.65$) is from Eq.\ (\ref{ft}).}
\label{fig04}
\end{figure}

The crossover frequency between the gravity and magneto-capillary regimes decreases when $B$ is increased (see Figure\ \ref{fig02}). Figure\ \ref{fig04} then shows the evolution of the rescaled crossover frequency $\tilde{f}(B)$ (see above) as a function of $B$ for 3 different frequency bandwidths of the random forcing. When $B$ is increased, $\tilde{f}$ is found to decrease with the same law whatever the forcing frequency. Beyond $B=B_c$, $\tilde{f}$ is roughly of the same order than the upper frequency of the forcing, and consequently can not be measured anymore. However, when $B > B_c$ and for the 1 - 6 Hz forcing, the power spectrum displays three power laws (not shown here) showing a magneto-capillary crossover. This new crossover frequency is reported in Fig.\ \ref{fig04} with ($\circ$)-symbols without rescaling. For lower frequency bandwidths of vibration, the slope breaking is too small to allow an accurate measurement of the transition between the magnetic and capillary regime.  

The evolutions of these crossover frequencies with $B$ are described as follows. Whatever $B$, Eq.\ (\ref{rdtheo}) is dominated, at small $k$, by the linear term (gravity waves) and, at high $k$, by the cubic term (capillary waves). One can assume that the quadratic term (magnetic waves) dominates when it is greater than the linear {\em and} cubic terms. This arises when $f[\chi]B^2 > \mu_0\sqrt{\rho g \gamma}$, that is, using Eqs.\ (\ref{aimantation}) and (\ref{BvsH}) and the ferrofluid properties, when $B>0.65B_c$. Thus, when $B<0.65B_c$, no wavelength exists for which the magnetic term dominates in Eq.\ (\ref{rdtheo}). When $B>0.65B_c$, magnetic waves exist on a range of wavelength between the gravity and capillary ones. This explains why both the spectrum exponent and its amplitude shown in Fig.\ \ref{fig03} change for $B\simeq 0.65B_c$. Surprisingly, this critical magnetic induction has never been reported previously. The crossover frequencies between the gravity, the magnetic and the capillary regimes are derived by balancing the dispersion relation terms each to each. For the gravity-capillary transition, one balances the first and the third terms of the second-hand of Eq.\ (\ref{rdtheo}), that is $gk_{gc}=(\gamma/\rho)k_{gc}^3$, thus for $k_{gc}=\sqrt{\rho g/\gamma}$ which substituted into Eq.\ (\ref{rdtheo}) gives
\begin{equation}
\omega^2_{gc}=2\sqrt{\frac{g^3\rho}{\gamma}}-\frac{gf[\chi]B^2}{\mu_0\gamma} {\rm \ , \ for \ \,Ê} f[\chi]B^2 < \mu_0\sqrt{\rho g \gamma} {\rm \ \ . }
\label{fgc}
\end{equation} 
Similarly, by balancing the first and second terms, the gravity-magnetic crossover frequency reads 
\begin{equation}
\omega^2_{gm}=\frac{\gamma}{\rho}\left[\frac{\mu_0\rho g}{f[\chi]}\right]^3B^{-6} {\rm \ , \ for \ \, } f[\chi]B^2 > \mu_0\sqrt{\rho g \gamma} {\rm \ \ . }
\label{fgm}
\end{equation} 
Finally, by balancing the second and the third terms, the magneto-capillary crossover frequency reads
\begin{equation}
\omega^2_{mc}=\frac{gf[\chi]}{\mu_0\gamma}B^2 {\rm \ , \ for  \ \, } f[\chi]B^2 > \mu_0\sqrt{\rho g \gamma} {\rm \ \ . }
\label{fmc}
\end{equation}
Using Eqs.\ (\ref{aimantation}) and (\ref{BvsH}) and the ferrofluid properties, these crossover frequency curves are plotted in Fig.\ \ref{fig04} as a function of $B/B_c$ and show a good agreement with the experimental data. This plot also shows the range of existence of magnetic waves for $B/B_c > 0.65$ and of a triple point corresponding to the coexistence of the three domains (gravity, magnetic and capillary ones). It can be derived by balancing the three terms of Eq.\ (\ref{rdtheo}), which leads to 
\begin{equation}
f_t=\frac{1}{2\pi}\left(\frac{g^3\rho}{\gamma}\right)^{1/4} {\rm \ \ and  \ } B_t^2 = \mu_0\sqrt{\rho g \gamma}/f[\chi(H_t)] {\rm \ \ , }
\label{ft}
\end{equation} 
corresponding to $f_t=10.8$ Hz and $B_t/B_c=0.65$ in good agreement with the data of Fig.\ \ref{fig04}.

\begin{figure}
\centerline{
\epsfxsize=75mm
\epsffile{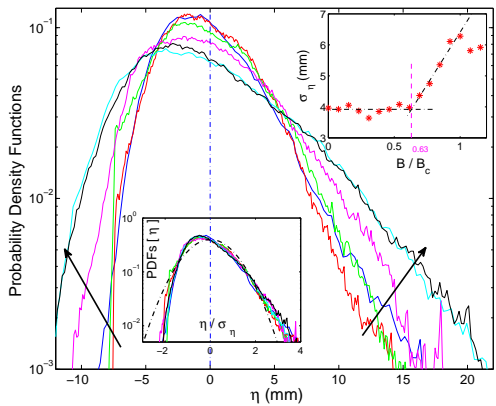} 
}
\caption{Probability density functions of wave amplitude, $\eta$, for different values of the dimensionless magnetic induction, from $B / B_c =$ 0, 0.3, 0.54, 0.77, 0.92 to 1.2 (see the arrows). Forcing parameter: $1 \leq f \leq 4$ Hz.  Bottom inset: Same PDFs displayed using the reduced variable $\eta/\sigma_{\eta}$. Gaussian fit (dashed line). Top inset: Standard deviation of wave amplitude $\sigma_{\eta}$ as a function of $B / B_c$}
\label{fig05}
\end{figure}

Finally, the probability density functions (PDFs) of the wave amplitudes are shown in Fig.\ \ref{fig05} for different value of $B/B_c$, at high enough amplitude of forcing. For $B=0$, the PDF is asymmetrical due to the strong steepness of the waves as for usual fluids \cite{Falcon07}. This means that the deep through are rare, whereas high crests are much more probable, thus showing the nonlinear nature of the wave interactions in a wave turbulence regime. The PDF asymmetry is enhanced when $B$ is increased. Note also that the most probable value of the PDF is more and more negative as $B$ increases, although its mean value $\langle \eta \rangle$ is zero as expected. The bottom inset of Fig.\ \ref{fig05} shows that all these PDFs normalized to its standard deviation value, $\sigma_{\eta}$, roughly collapse on a single non Gaussian distribution.  This means that the distribution depends only on $\sigma_{\eta}$. The top inset of Fig.\ \ref{fig05} then shows the evolution of $\sigma_{\eta}$ as a function of $B/B_c$. For $B/B_c \leq 0.63$, the rms value amplitude of the waves does not depend on the magnetic induction.  Note that this value is very close to the above predicted onset of magnetic waves $B_t/B_c=0.65$. When this onset is exceed,  $\sigma_{\eta}$ increases roughly linearly with $B$ up to the occurrence of the Rosensweig instability at $B_c$.

\begin{acknowledgments}
We thank D. Talbot for the synthesis of the ferrofluid, J.-C. Bacri and A. Cebers for fruitfull discussion, A. Lantheaume, and C. Laroche for technical assistance. This work has been supported by ANR Turbonde BLAN07-3-197846.
\end{acknowledgments}

\end{document}